\documentstyle[11pt]{article}
\newcommand{\be}[1]{\begin{equation}\label{#1}}
\newcommand{\ba}[1]{\begin{eqnarray}\label{#1}}
\newcommand{\ee}{\end{equation}}
\newcommand{\ea}{\end{eqnarray}}
\newcommand{\non}{\nonumber\\\rule{0pt}{30pt}}
\newcommand{\nona}[1]{\nonumber\\\rule{0pt}{#1pt}}
\newcommand{\num}{\\\rule{0pt}{20pt}}
\newcommand{\numa}[1]{\\\rule{0pt}{#1pt}}
\newcommand{\dis}{\displaystyle}
\renewcommand{\Im}{\mathop{\rm Im}}
\newcommand{\TR}{\mathop{\rm Tr}}
\newcommand{\Tr}{\mathop{\rm tr}}
\newcommand{\mm}{\mu}

\newcommand{\freop}{\Bigl(\tilde I+\widetilde V\Bigr)}

\newcommand{\EP}[1]{E_+(#1|u)}
\newcommand{\EL}{\langle E^L(\lambda)|}
\newcommand{\ER}{|E^R(\mu)\rangle}
\newcommand{\ELm}{\langle E^L(\mu)|}
\newcommand{\ERl}{|E^R(\lambda)\rangle}
\newcommand{\ERls}{E^R(\lambda)\rangle}

\newcommand{\Elv}[1]{E_{#1}^L(\lambda|v)}

\newcommand{\FL}{\langle F^L(\lambda)|}
\newcommand{\FR}{|F^R(\mu)\rangle}
\newcommand{\FLm}{\langle F^L(\mu)|}
\newcommand{\FRl}{|F^R(\lambda)\rangle}

\newcommand{\Fl}[1]{F_{#1}^L(\lambda|u)}

\newcommand{\Fr}[1]{F_{#1}^R(\mu|u)}
\newcommand{\Frl}[1]{F_{#1}^R(\lambda|u)}

\newcommand{\EM}[1]{E_-(#1|u)}

\newcommand{\tV}{\widetilde V}

\newcommand{\hs}{\hat\sigma}
\newcommand{\hb}{\hat b}
\newcommand{\hB}{\widehat B}
\newcommand{\hC}{\widehat C}

\newcommand{\hg}{\hat g}
\newcommand{\hG}{\widehat G(\lambda)}
\newcommand{\hGm}{\widehat G^{(m)}}
\newcommand{\hGe}{\widehat G^{(e)}}

\newcommand{\hx}{\hat\chi}
\newcommand{\hmod}{\hat\chi^{(m)}}
\newcommand{\hxl}{\hat\chi(\lambda)}
\newcommand{\hxp}{\hat\chi_+(\lambda)}
\newcommand{\hxpm}{\hat\chi_\pm(\lambda)}
\newcommand{\hxm}{\hat\chi_-(\lambda)}
\newcommand{\ketd}{|0)}
\newcommand{\brad}{(0|}
\newcommand{\Eq}[1]{(\ref{#1})}
\newcommand{\stint}{\int\limits_{-\infty}^\infty}
\newcommand{\qint}{\int\limits_{-\infty}^\infty}

\newcommand{\RH}{Riemann--Hilbert problem\ }
\newcommand{\rh}{Riemann--Hilbert problem}

\newlength{\minitwocolumn}
\catcode`\@=11
\def\relaxnext@{\let\next\relax}
\def\eq#1\endeq{\begin{eqnarray}#1\end{eqnarray}}
\def\eqn#1\endeqn{\begin{eqnarray*}#1\end{eqnarray*}}

\def\eq#1\endeq{\begin{eqnarray}#1\end{eqnarray}}
\def\eqn#1\endeqn{\begin{eqnarray*}#1\end{eqnarray*}}

\makeatletter
\@addtoreset{equation}{section}
\makeatother

\newtheorem{thm}{Theorem}[section]

\begin{document}
%%%%%%%%%%%%%%%%%%%%%%%%%%%%%%%%%%%%%%%%%%%%%%%%%%%%%%%%%%
\begin{flushright}
June\\
1997
\end{flushright}
\vspace{24pt}
\begin{center}
\begin{Large}
{\bf The Riemann--Hilbert problem associated
with the quantum Nonlinear Schr\"odinger equation.}
\end{Large}

\vspace{26pt}

\begin{large}
V.~E.~Korepin\raisebox{2mm}{{\scriptsize a}}
\end{large}

\begin{small}
{\it Institute for Theoretical Physics, State University of
New York at Stony Brook,
Stony Brook, NY 11794-3840, USA.}
\end{small}

\begin{large}
N.~A.~Slavnov\raisebox{2mm}{{\scriptsize b}}
\end{large}

\begin{small}
{\it Steklov Mathematical Institute,
Gubkina 8 , Moscow 117966, Russia.}
\end{small}
\vspace{22pt}

\underline{Abstract}
\end{center}
\vspace{14pt}

We consider the dynamical correlation functions of the quantum Nonlinear
Schr\"odinger equation.  In a previous paper \cite{KKS1} we found that
the dynamical correlation functions can be described by the vacuum
expectation value of an operator-valued Fredholm determinant.  In this
paper we show that a Riemann--Hilbert problem can be associated with this
Fredholm determinant.  This Riemann--Hilbert problem formulation permits
us to  write down completely integrable equations for the Fredholm
determinant and to perform  an asymptotic analysis for the
correlation function.

\vskip 3mm

\vfill
\hrule

\begin{footnotesize}
\noindent
\raisebox{2mm}{$a$}
korepin@insti.physics.sunysb.edu,~\
\raisebox{2mm}{$b$}
nslavnov@mi.ras.ru
\end{footnotesize}
\newpage
%%%%%%%%%%%%%%%%%%%%%%%%%%%%%%%%%%%%%%%%%%%%%%%%%%%%%%%%%%%
\section{Introduction}
%%%%%%%%%%%%%%%%%%%%%%%%%%%%%%%%%%%%%%%%%%%%%%%%%%%%%%%%%%%

We consider exactly solvable models of statistical mechanics in one
space and one time dimension. The Quantum Inverse Scattering Method
and the Algebraic Bethe Ansatz are effective methods for a
description of the spectrum of these models. Our aim is the
evaluation of correlation functions of exactly solvable models. Our
approach is based on the determinant representation for correlation
functions. It consists of a few steps:  first the correlation
function is represented as a vacuum expectation-value
of an operator-valued Fredholm determinant, second  the Fredholm
determinant is described by a classical completely integrable
equation, third the classical completely integrable equation is
solved by means of the Riemann--Hilbert problem, fourth, the vacuum
expectation-value of the asymptotics for the operator-valued Fredholm
determinant is calculated.  This permits us to evaluate the long
distance and large time asymptotics of the correlation function.  Our
method is described in \cite{KBI}.

The  quantum Nonlinear Schr\"odinger equation  can be
described in terms of the canonical Bose fields
$\psi(x,t),~\psi^{\dagger}(x,t),~(x,t \in {\bf R})$
with the standard commutation relations
\begin{eqnarray}
[\psi(x,t), \psi^{\dagger}(y,t)]=\delta(x-y),\hspace{2pt}
[\psi(x,t), \psi(y,t)]=[\psi^{\dagger}(x,t), \psi^{\dagger}(y,t)]=0.
\end{eqnarray}
The Hamiltonian of the model is
\be{Hamilton}
{H}={\dis\int \,dx
\left({\partial_x}\psi^{\dagger}(x)
{\partial_x} \psi(x)+
c\psi^{\dagger}(x)\psi^{\dagger}(x)\psi(x)\psi(x)
-h \psi^{\dagger}(x)\psi(x)\right).}
\ee
Here $0<c<\infty$ is the coupling constant and $h>0$
is the chemical potential. The Hamiltonian $H$ acts in the Fock space
with the vacuum vector $|0\rangle$.  The vacuum vector
$|0\rangle$ is characterized by the relation:
\begin{eqnarray}
\psi(x,t)|0\rangle =0.
\end{eqnarray}
The  vacuum vector $\langle0|$ is characterized by
the relation:
\begin{eqnarray}
\langle 0| \psi^\dagger(x,t)=0,~
\langle0|0\rangle=1.
\end{eqnarray}
The spectrum of the model was first described by E.~H.~Lieb and
 W.~Liniger \cite{L}, \cite{LL}.
The Lax representation for the corresponding classical equation of
motion
\begin{eqnarray}
i\frac{\partial}{\partial t} \psi=
[\psi, H]=
-\frac{\partial^2}{{\partial x}^2} \psi
+2c \psi^{\dagger}\psi \psi
-h \psi,
\end{eqnarray}
was found by V.~E.~Zakharov and A.~B.~Shabat \cite{ZS}.
The Quantum Inverse Scattering Method for the model
was formulated by L.~D.~Faddeev and E.~K.~Sklyanin  \cite{FS}.

In this paper we shall consider the thermodynamics of the model.
The partition function ${\cal Z}$ and free energy $F$ are defined by
\begin{eqnarray}
{\cal Z}=\Tr\left(e^{-\frac{H}{T}}\right)=e^{-\frac{F}{T}}.
\end{eqnarray}
The free energy $F$ can be expressed in terms of the Yang--Yang 
equation \cite{YY}
\ba{YY} 
{\dis\varepsilon(\lambda)}&=&{\dis\lambda^2-h-\frac{T}{2\pi}
\int_{-\infty}^{\infty}
\frac{2c}{c^2+(\lambda-\mu)^2}\ln
\left(1+e^{-\frac{\varepsilon(\mu)}{T}}
\right)d\mu,}\num
{\dis F}&=&{\dis-\frac{T}{2 \pi} \int_{-\infty}^{\infty}\ln\left(
1+e^{-\frac{\varepsilon(\mu)}{T}}\right)d \mu.}\label{Freeen}
\ea
The correlation function, which we shall study in this paper, is
defined by
\be{tempcorrel}
\langle\psi(0,0)\psi^\dagger(x,t)\rangle_T=
\frac{\Tr\left( e^{-\frac HT}\psi(0,0)\psi^\dagger(x,t)\right)
}
{\Tr e^{-\frac HT}}.
\ee
In a previous paper \cite{KKS1} we obtained the determinant
representation for this correlation function. It was shown in the
paper \cite{KKS2} that the Fredholm determinant thus obtained can be
expressed in terms of solutions of an non-Abelian nonlinear 
Schr\"odinger equation. More precisely, the second logarithmic 
derivatives of the Fredholm determinant with respect to distance and 
time are densities of the conservation laws of this equation. 
However, such a description of the determinant is not complete, since 
the non-Abelian Nonlinear Schr\"odinger equation has an infinite set 
of solutions, and it is not clear a priori, which of them describes 
the correlation function.

On the contrary, the  Riemann--Hilbert problem associated with
the Fredholm determinant is uniquely solvable. This is the main advantage
of the approach based the  application of the Riemann--Hilbert
problem.

The plan of this paper is the following.
In section 2 we  review the determinant representation and recall
definitions and notations used in \cite{KKS1}, \cite{KKS2}.
In section 3 we formulate the Riemann--Hilbert problem. We prove
the equivalence of the integral equations considered in \cite{KKS2} and
the Riemann--Hilbert problem.
The Lax representation of the  non-Abelian  Nonlinear Schr\"odinger
equation is considered in section 4. Section 5 is devoted to  the 
modified formulation of the Riemann--Hilbert problem, which is 
especially useful for asymptotic analysis.  
% 
%%%%%%%%%%%%%%%%%%%%%%%%%%%%%%%%%%%%%%%%%%%%%%%%%%%%%%%%%%
\section{Determinant representation for the correlation function}
%%%%%%%%%%%%%%%%%%%%%%%%%%%%%%%%%%%%%%%%%%%%%%%%%%%%%%%%%%%

In this section we summarize the formulations and results
obtained in the previous papers \cite{KKS1}, \cite{KKS2}
for the reader's convenience. Our starting point is the determinant
representation for the dynamical correlation function of the local fields
obtained in \cite{KKS1}:
\begin{eqnarray} \langle
\psi(0,0)\psi^\dagger(x,t)\rangle_T= -\frac{e^{-iht}}{2 \pi}
\brad
\frac{\det\left(\tilde{I}+\widetilde{V}\right)}
{\det\left(\tilde{I}-\frac{1}{2\pi}{\widetilde K}_T\right)}
\displaystyle \int_{-\infty}^{\infty}
\hb_{12}(u,v)du dv \ketd.\label{detrep}
\end{eqnarray}
In order to describe explicitly the r.h.s. of \Eq{detrep} we present
here the system of definitions and notations, used in \cite{KKS2}.
%%%%%%%%%%%%%%%%%%%%%%%%%%%%%%%%%%%%%%%%%%%%%%%%%%%%%%%%%%%%%%
\subsection{Vectors and operators of the Hilbert space}
%%%%%%%%%%%%%%%%%%%%%%%%%%%%%%%%%%%%%%%%%%%%%%%%%%%%%%%%%%%%%%

Consider ket-vector $\ERl$
\be{eright0}
\ERl=\left(\begin{array}{c}E^R_1(\lambda \vert u)\\
E^R_2(\lambda \vert u)\end{array}
\right),
\ee
belonging to a Hilbert space $\cal H$. Here $E_j^R(\lambda|u)$ are
two-variable functions. It is convenient to present the space $\cal
H$ as the tensor product of two spaces ${\cal H}=\widetilde{\cal H}
\otimes\widehat{\cal H}$. Each of the discrete components
$E_j^R(\lambda|u)$, being a function of the variable $\lambda$, belongs
to the space $\widetilde{\cal H}$. In turn, the space $\widehat{\cal H}$
consists of two-component functions of the  variable $u$.

In order to define scalar products in the space $\widehat{\cal H}$
we introduce the bra-vector $\EL$
\be{eleft0}
\langle E^L(\lambda)\vert=
\left(E_1^L(\lambda \vert u), E_2^L(\lambda \vert u)
\right),
\ee
which also can be treated as vector of $\cal H$. Then the scalar
product in $\widehat{\cal H}$ is given by standard formula
\be{defscalprod}
\langle E^L(\lambda)\vert E^R(\mu)\rangle=
\stint \left(E_1^L(\lambda|u)E_1^R(\mu|u)+
E_2^L(\lambda|u)E_2^R(\mu|u)\right)\,du.
\ee

Below we shall consider two types of operators acting in $\cal H$.
The first type of operators of which  $\tilde I+
\widetilde V$ in \Eq{detrep} is an example, acts in the space 
$\widetilde{\cal H}$
only. By definition
\be{actiontilde}
\freop \circ \vert E^R(\mu)\rangle
=\left(\begin{array}{c}\displaystyle E^R_1(\lambda\vert u)+
\stint \widetilde V(\lambda,\mu)E^R_1(\mu\vert u) d\mu \numa{33}
{\displaystyle E^R_2(\lambda\vert u)+
\stint \widetilde V(\lambda,\mu)E^R_2(\mu\vert u) d\mu}
\end{array}\right),
\ee
\be{actiontilde1}
\ERl\circ\freop
=\left(\begin{array}{c}\displaystyle E^R_1(\mu\vert u)+
\stint  E^R_1(\lambda\vert u)\widetilde V(\lambda,\mu) d\lambda
 \numa{33}
\displaystyle E^R_2(\mu\vert u)+
\stint  E^R_2(\lambda\vert u)\widetilde V(\lambda,\mu) d\lambda
\end{array}\right).
\ee
Here $\tilde I$ is the identity operator in $\widetilde{\cal H}$. The
action on the bra-vectors $\EL$ is quite similar. Below we shall mark
the operators of this type by ``tilde" and denote their action by the
sign ``$\circ$".

The second type of operator acts in the space $\widehat{\cal H}$. It
is convenient to treat these operators as $2\times 2$ matrices with
operator-valued entries:
\begin{eqnarray}
\hat A=\left(
\begin{array}{cc}
{\dis \hat A_{11}(u,v)} &{\dis  \hat A_{12}(u,v)}\num
{\dis  \hat A_{21}(u,v)} &{\dis   \hat A_{22}(u,v)}
\end{array}\right).\label{hatoperator}
\end{eqnarray}
The action of the operators \Eq{hatoperator} is given by
\be{action}
\hat{A} \cdot \vert
E^R(\lambda)\rangle = \left(\begin{array}{c}\displaystyle
\sum_{k=1}^2 \stint \hat A_{1k}(u,v)
E^R_k(\lambda \vert v)\, dv \nona{30}
\displaystyle
\sum_{k=1}^2 \stint \hat A_{2k}(u,v) E^R_k(\lambda \vert v) \,dv
\end{array}\right).
\ee
\be{action1}
{\dis \EL\cdot \hat A=\left(
\sum_{j=1}^{2}\stint E^L_j(\lambda|u)\hat A_{j1}(u,v)\,du;
\sum_{j=1}^{2}\stint E^L_j(\lambda|u)\hat A_{j2}(u,v)\,du\right)}
\ee
This type of operator will be marked by ``hat".

We would like to draw to the  attention of the reader that we use ``hat" 
not only for the operator-valued matrices of type \Eq{hatoperator}, 
but for their matrix elements also. We hope, that this system of 
notation will not cause misunderstanding.

Finally, we define the trace of ``hat"-operators as
\be{Trace}
\TR\hat A=\Tr\hat A_{11}+\Tr\hat A_{22}=
\stint\left(\hat A_{11}(u,u)+\hat A_{22}(u,u)\right)\,du,
\ee
where
\be{trace}
\Tr\hat A_{jk}=\stint\hat A_{jk}(u,u)\,du.
\ee
%
%%%%%%%%%%%%%%%%%%%%%%%%%%%%%%%%%%%%%%%%%%%%%%%%%%%%%%%%%%
\subsection{Dual fields}
%%%%%%%%%%%%%%%%%%%%%%%%%%%%%%%%%%%%%%%%%%%%%%%%%%%%%%%%%%

The important objects, on which the r.h.s. of \Eq{detrep} depends,
are auxiliary quantum operators --- dual fields, acting in an
auxiliary Fock space.  They are introduced in order to remove two-body
scattering and to reduce the model to free fermionic. One can find the
detailed definition and properties of dual fields in Section 5 and 
Appendix C of \cite{KKS1}.  Here we repeat them in brief.

Consider an auxiliary Fock space having vacuum vector $\ketd$ and
dual vector $\brad$. The three dual fields $\psi(\lambda)$,
$\phi_{D_1}(\lambda)$ and $\phi_{A_2}(\lambda)$ acting in this space
are defined as
\be{dualfields}
\begin{array}{rcl}
\phi_{A_2}(\lambda)&=&q_{A_2}(\lambda)+p_{D_2}(\lambda),\\
\phi_{D_1}(\lambda)&=&q_{D_1}(\lambda)+p_{A_1}(\lambda),\\
\psi(\lambda)&=&q_\psi(\lambda)+p_\psi(\lambda).
\end{array}
\ee
Here $p(\lambda)$ are annihilation parts of dual fields:
$p(\lambda)\ketd=0$; $q(\lambda)$ are creation parts of dual fields:
$\brad q(\lambda)=0$.  Thus, any dual field is the sum of
annihilation and creation parts (the dual field $\psi(\lambda)$
should not be confused with the field $\psi(x,t)$, which appears
in the expression for the Hamiltonian of the model).

The only nonzero commutation
relations are
\be{commutators}
\begin{array}{l}
{}[p_{A_1}(\lambda),q_\psi(\mu)]=
[p_\psi(\lambda),q_{A_2}(\mu)]=\ln h(\mu,\lambda),\num
{}[p_{D_2}(\lambda),q_\psi(\mu)]=
[p_\psi(\lambda),q_{D_1}(\mu)]=\ln h(\lambda,\mu),\num
{}[p_\psi(\lambda),q_\psi(\mu)]=\ln [h(\lambda,\mu)h(\mu,\lambda)],
\quad\mbox{where}\quad
 {\dis h(\lambda,\mu)=\frac{\lambda-\mu+ic}{ic}.}
\end{array}
\ee
Recall that $c$ is the coupling constant in \Eq{Hamilton}.
It follows immediately from \Eq{commutators} that the dual fields
belong to an Abelian sub-algebra
\be{Abel} [\psi(\lambda),\psi(\mu)]=
[\psi(\lambda),\phi_a(\mu)]=
[\phi_b(\lambda),\phi_a(\mu)]=0,
\ee
where $a,b=A_2,D_1$. The properties \Eq{commutators}, \Eq{Abel}, in fact,
permit us to treat the dual fields as complex functions,
which are holomorphic in some neighborhood of the real axis.
%
%%%%%%%%%%%%%%%%%%%%%%%%%%%%%%%%%%%%%%%%%%%%%%%%%%%%%%%%%%%%%%%
\subsection{Fredholm determinant}
%%%%%%%%%%%%%%%%%%%%%%%%%%%%%%%%%%%%%%%%%%%%%%%%%%%%%%%%%%%%%%%

Now we are ready to describe the determinant representation
\Eq{detrep}.

The important factor entering the r.h.s. of (\ref{detrep}) is the
Fredholm determinant of the integral operator $\tilde I+\tV$,
acting on the real axis as
\begin{eqnarray}
\left(\tilde{I}+\widetilde{V}\right)\circ f(\mu)
=f(\lambda)+\displaystyle \int_{-\infty}^{\infty}\tV(\lambda,\mu)
f(\mu)d\mu,
\end{eqnarray}
 where $f(\lambda)$ is some trial function.
Recall that $\tilde I$ is the identity operator in the space
$\widetilde{\cal H}$.

The kernel $\tV(\lambda,\mu)$ can be presented in the form
\be{kernel01}
\tV(\lambda,\mu)=\frac{\langle E^L(\lambda)\vert E^R(\mu)\rangle}
{\lambda-\mu},
\ee
where the components of vectors $\langle E^L(\lambda)\vert$ and $\vert
E^R(\mu)\rangle$ are equal to
\ba{eplus0}
&&{\dis
E_1^R(\lambda|u)=E_2^L(\lambda|u)=E_+(\lambda|u),}\numa{20}
&&{\dis
E_2^R(\lambda|u)=-E_1^L(\lambda|u)=E_-(\lambda|u).}
\label{eminus0}
\ea
Due to this specification we have
\be{ortho}
\EL\ERls=0,
\ee
hence  the kernel $\tV(\lambda,\mu)$ possesses no singularity at
$\lambda=\mu$. Later on, the orthogonality property \Eq{ortho}
will play an important role.

The functions $E_\pm$ introduced in \cite{KKS2} are equal to:
\ba{eplus}
&&{\dis
\EP\lambda=\frac{1}{2\pi}\frac{Z(u,\lambda)}{Z(u,u)}
\left(\frac{e^{-\phi_{A_2}(u)}}{u-\lambda+i0}
+\frac{e^{-\phi_{D_1}(u)}}{u-\lambda-i0}\right)
\sqrt{\vartheta(\lambda)}}\nona{19}
&&\hskip3cm{\dis \times
e^{\psi(u)+\tau(u)+\frac12(
\phi_{D_1}(\lambda)+\phi_{A_2}(\lambda)-\psi(\lambda)-\tau(\lambda))},}
\ea
\hskip1mm
\be{eminus}
\EM\lambda=\frac{1}{2\pi}Z(u,\lambda)
e^{\frac12(\phi_{D_1}(\lambda)+\phi_{A_2}(\lambda)
-\psi(\lambda)-\tau(\lambda))}\sqrt{\vartheta(\lambda)},
\ee
where the function $Z(\lambda,\mu)$ is defined by
\be{Z}
Z(\lambda,\mm)=\frac{e^{-\phi_{D_1}(\lambda)}}{h(\mm,\lambda)}+
\frac{e^{-\phi_{A_2}(\lambda)}}{h(\lambda,\mm)}.
\ee
Here $\psi(\lambda)$, $\phi_{D_1}(\lambda)$ and $\phi_{A_2}(\lambda)$
are just the dual fields \Eq{dualfields}.

Recall also that the function $\vartheta(\lambda)$ is the Fermi weight
\be{Fermi}
\vartheta(\lambda)=\left(1+\exp\left[
\frac{\varepsilon(\lambda)}{T}\right]\right)^{-1}.
\ee
This function defines the dependence of the correlation function on
temperature and chemical potential. The function $\tau(\lambda)$ is the only
function depending on time and distance:
\be{tau}
\tau(\lambda)=it\lambda^2-ix\lambda.
\ee
Thus, functions $E_\pm(\lambda|u)$ depend on time $t$, distance $x$,
temperature $T$ and chemical potential $h$, but this dependence, as a
rule, suppressed in our notations.

In order to define the function $\hb_{12}(u,v)$, we
introduce  vectors $\FL$ and $\FR$, belonging to the space
$\cal H$
\be{introF} \FL=\Bigl(\Fl1,\Fl2\Bigr);\qquad\qquad
\FR=\left(\begin{array}{c}
\Fr1\\
{}\\
\Fr2
\end{array}\right),
\ee
as solutions of  the integral equations (see \cite{KKS2})
\ba{intequation1}
&&\hspace{-13mm}\freop\circ\FLm
\equiv\FL+\stint\tV(\lambda,\mu)\FLm\,d\mu
=\EL,\\
&&\hspace{-13mm}\FRl\circ\freop
\equiv\FR+\stint\FRl\tV(\lambda,\mu)\,d\lambda
=\ER.
\label{intequation2}
\ea

Let us define operator  $\hB$ as
\be{B1}
\hB=\stint\FRl\EL\,d\lambda
=\stint\ER\FLm\,d\mu.
\ee
The  components of this operator are
\be{componentB}
B_{jk}(u,v)=\stint\Frl{j}\Elv{k}\,d\lambda,
\qquad j,k=1,2.
\ee
The function $\hb_{12}(u,v)$ is the matrix element of
the operator $\hat{b}$ defined by
\begin{eqnarray}
\hb=\hB+\hg.\label{defb}
\end{eqnarray}
Here  $\hat{g}$ is
\begin{eqnarray}
\hat{g}=\left(\begin{array}{cc}0 & -1
\\0 &0\end{array}\right)\hg_{12}(u,v),
\qquad
\hg_{12}(u,v)=\delta(u-v)e^{\psi(v)+\tau(v)}.\label{defg}
\end{eqnarray}
Due to the above definition we have
\begin{eqnarray}
\hb_{ab}(u,v)&=&\hB_{ab}(u,v)
\qquad\mbox{for all $a,b$ except  $a=1,b=2$,}\\
\hb_{12}(u,v)&=&\hB_{12}(u,v)-\hg_{12}(u,v)
\end{eqnarray}
The importance of operators $\hb$ and $\hg$ will be clear later, after
the formulation of the Riemann--Hilbert problem.

The last factor in the representation of the correlation function,
which is not explained yet, is the determinant of the operator
$\tilde I-\frac{1}{2\pi}{\widetilde K}_T$. This operator  also acts on
the whole real axis. Its kernel is given by
\be{KT}
\widetilde K_T(\lambda,\mu)=\left(\frac{2c}{c^2+
(\lambda-\mu)^2}\right)
\sqrt{\vartheta(\lambda)}\sqrt{\vartheta(\mu)}.
\ee
This determinant does not depend on time and distance, therefore
it can be considered as a constant factor.

Thus, we have described the r.h.s. of (\ref{detrep}). The temperature
correlation function of local fields is proportional to the
vacuum expectation value in  the auxiliary Fock space of the Fredholm
determinant of the integral operator. The  auxiliary quantum
operators---dual fields---enter the kernels $\tV$ and $\hb_{12}(u,v)$.
However, due to the property \Eq{Abel} the Fredholm determinant is
well defined.
%
%%%%%%%%%%%%%%%%%%%%%%%%%%%%%%%%%%%%%%%%%%%%%%
\section{Riemann--Hilbert problem}
%%%%%%%%%%%%%%%%%%%%%%%%%%%%%%%%%%%%%%%%%%%%%%

Consider the Riemann--Hilbert problem for the operator
$\hat{\chi}(\lambda)$ acting in the space $\widehat {\cal H}$ and
depending on the complex parameter $\lambda$:
\begin{eqnarray}
\hat\chi(\lambda \vert u,v)=\left(\begin{array}{cc}
\hat\chi_{11}(\lambda \vert u,v) & \hat\chi_{12}(\lambda \vert u,v)\\
\hat\chi_{21}(\lambda \vert u,v) & \hat\chi_{22}(\lambda \vert u,v)
\end{array}\right).
\end{eqnarray}
The matrix $\hat\chi(\lambda\vert u,v)$ is holomorphic for $\lambda \in
{\bf C} \setminus {\bf R}$, and satisfies the normalization condition at
$\lambda=\infty$
\begin{eqnarray}
\hat\chi(\infty \vert
u,v)=\hat I\equiv\left(\begin{array}{cc}1 & 0 \\ 0
&1\end{array}\right)\delta(u-v).\label{eqn:inf}
\end{eqnarray}
The boundary values on the real axis are related by
\begin{eqnarray}
\hat{\chi}_{-}(\lambda)=
\hat{\chi}_{+}(\lambda)\hat{G}(\lambda),~~\lambda \in {\bf R},
\label{RH1}
\end{eqnarray}
where we set
$
\hat{\chi}_{\pm}(\lambda)=\lim_{\epsilon \to +0}
\hat{\chi}(\lambda \pm i \epsilon).
$
The jump matrix $\hat{G}(\lambda)$ here is given by
\begin{eqnarray}\label{hG}
\hat G(\lambda \vert u,v)=\left(\begin{array}{cc} 1 & 0 \\
0 & 1 \end{array}\right)\delta(u-v)+2\pi i \hat H(\lambda \vert u,v),
\end{eqnarray}
where
\begin{equation}\label{hH}
\hat H(\lambda \vert u,v)=\vert E^R(\lambda)\rangle
\langle E^L(\lambda)\vert
=\left(\begin{array}{cc}
E_1^R(\lambda \vert u)E_1^L(\lambda \vert v) &
E_1^R(\lambda \vert u)E_2^L(\lambda \vert v) \num
E_2^R(\lambda \vert u)E_1^L(\lambda \vert v) &
E_2^R(\lambda \vert u)E_2^L(\lambda \vert v)
\end{array}\right).
\end{equation}
Recall that due to \Eq{eplus0}, \Eq{eminus0}
\begin{eqnarray}
\begin{array}{cc}
E_1^R(\lambda \vert u)=E_+(\lambda \vert u),&
E_2^R(\lambda \vert u)=E_-(\lambda \vert u),\num
E_1^L(\lambda \vert v)=-E_-(\lambda \vert v),&
E_2^L(\lambda \vert v)=E_+(\lambda \vert v),
\end{array}\label{str1}
\end{eqnarray}
where the functions $E_{\pm}(\lambda \vert u)$ are defined in
(\ref{eplus}) and (\ref{eminus}). However, we do not use the explicit
expressions for $E_\pm$ in this section.

One should understand the r.h.s of the jump condition \Eq{RH1} as an
 operator product in the space $\widehat{\cal H}$. Another
words, more detailed the equality \Eq{RH1}  means
\begin{eqnarray}
\Bigl(\hat{\chi}_{-}(\lambda|u,v)\Bigr)_{jk}=
\sum_{l=1}^{2}\stint
\Bigl(\hat{\chi}_{+}(\lambda|u,w))\Bigr)_{jl}
\Bigr(\hat{G}(\lambda|w,v)\Bigr)_{lk}\,dw.
\label{RH11}
\end{eqnarray}
Thus, we are dealing with an infinite dimensional \rh.

The orthogonality property \Eq{ortho}
\begin{eqnarray}
\langle E^L(\lambda)\vert E^R(\lambda)\rangle=0\label{eqn:zero}
\end{eqnarray}
implies $\TR \hat{H}^n(\lambda)=0$  for  $ ~~n\ge1$ .  Hence,
\begin{eqnarray}
\det \hat{G}(\lambda)=1.\label{eqn:detG=1}
\end{eqnarray}

We assume that the Riemann--Hilbert problem is solvable. We note also
that in the class of integral operators which we are 
 dealing with, the analyticity of the operator with respect
to the parameter $\lambda$ implies the analyticity of its
determinant. Therefore, due to the Liouville theorem and equations
(\ref{eqn:inf}), (\ref{eqn:detG=1}), we conclude that $ \det
\hat{\chi}(\lambda)=1$.  Hence, as in  to the usual matrix case the
solvability of our \RH implies the uniqueness of the solution.  It
means also that the inverse matrix $\hat\chi^{-1}(\lambda)$
exists and  is holomorphic for $\lambda \in {\bf C}
\setminus {\bf R}$.

The integral equations \Eq{intequation1}, \Eq{intequation2}
\ba{intequation1p}
&&\freop\circ\FLm=\EL,\\
&&\FRl\circ\freop=\ER,
\label{intequation2p}
\ea
were used in \cite{KKS2} for derivation of the differential equation,
describing the Fredholm determinant.
Now we prove the equivalence of these integral equations and the
Riemann--Hilbert problem \Eq{eqn:inf}, \Eq{RH1}.
Namely, we establish that the solution of the
Riemann--Hilbert problem can be presented in terms of the solutions
of integral equations and vice versa.

The following integral representation is the basis of this paper.
\begin{thm}~~~
A solution of the Riemann--Hilbert problem
(\ref{RH1}) has the following integral
representation
\be{solutchi1}
\hx(\lambda)=\hat I-\qint\frac{\FR\ELm}
{\mu-\lambda}\,d\mu,
\ee
where $\FR$ is the solution of the integral equation
(\ref{intequation2p}):
$$
\FRl\circ\freop=\ER.
$$
The inverse of this solution has the following integral representation:
\be{solutchi2}
\hx^{-1}(\lambda)=\hat I+\qint\frac{\ER\FLm}
{\mu-\lambda}\,d\mu,
\ee
where $\FL$ is the solution of integral equation
\Eq{intequation1p}:
$$
\freop \circ \langle F^L(\mu)\vert=\EL.
$$
\end{thm}
{\sl Proof.}~~
The operator $\hx(\lambda)$ defined in (\ref{solutchi1}) is
a holomorphic function everywhere except the real axis
and possesses the correct asymptotic behavior when $\lambda
\to\infty$. On the other hand, on the real axis we have
\ba{proofsolution}
{\dis
\hxp\hG}&=&{\dis \hxp+2\pi i\Biggl[\ERl\EL}\non
&-&{\dis
\qint\frac{\FR\ELm\ERls\EL}{\mu-\lambda-i0}\,d\mu\Biggr]}\non
&=&{\dis
\hxp+2\pi i\left[\ERl-\qint
\FR \tV(\mu,\lambda)\,d\mu\right]\EL}\non
&=&{\dis
\hat I-\qint\frac{\FR\ELm}
{\mu-\lambda-i0}\,d\mu+2\pi i\FRl\EL}\non
&=&{\dis \hat I-\qint\frac{\FR\ELm}
{\mu-\lambda+i0}\,d\mu=\hxm.}
\ea
Thus, the operator-valued matrix $\hx(\lambda)$ satisfies all the
conditions of the \RH \Eq{eqn:inf}, \Eq{RH1}.

Just the same method allows one to check representation \Eq{solutchi2}.
The theorem is proved.

A direct corollary of the representation \Eq{solutchi1} is the
following asymptotic expansion of $\hat{\chi}_{\pm}(\lambda)$ at
$\lambda\to\infty$
\be{expansion1}
\hxpm=\hat
I+\frac{\hB}{\lambda}+\frac{\hC}{\lambda^2}+\dots,
\ee
where
\be{deccoeff}
\hB=\stint\FRl\EL\,d\lambda,\quad
\hC=\stint \lambda\FRl\EL\,d\lambda, \qquad\dots
\ee
We see that the operator $\hB$ introduced in the previous section
(see \Eq{B1}) appears as  the first coefficient in the  asymptotic
expansion \Eq{expansion1}.

\begin{thm}~~~For arbitrary $\lambda\in \bf{C}$
\begin{eqnarray}
|F^R(\lambda)\rangle&=&\hat\chi(\lambda)|E^R(\lambda)\rangle,
\label{gauge1}\\
\langle F^L(\lambda)|&=&\langle E^L(\lambda)|\hat\chi^{-1}(\lambda).
\label{gauge2}
\end{eqnarray}
In particular, for  $\lambda\in {\bf R}$
\begin{eqnarray}
|F^R(\lambda)\rangle&=&\hat\chi_+(\lambda)|E^R(\lambda)\rangle
=\hat\chi_-(\lambda)|E^R(\lambda)\rangle,
\label{gauge1'}\\
\langle F^L(\lambda)|&=&\langle E^L(\lambda) |\hat\chi^{-1}_+(\lambda)
=\langle E^L(\lambda)|\hat\chi^{-1}_-(\lambda).\label{gauge2'}
\end{eqnarray}
\end{thm}
{\sl Proof.}~~
Using the definition of vectors $|F^R(\lambda)\rangle$ and
$\langle F^L(\lambda)|$ and integral representations for
$\hat\chi(\lambda)$ and $\hat\chi^{-1}(\lambda)$ we have
\begin{eqnarray}
\hat\chi(\lambda)|E^R(\lambda)\rangle&=&|E^R(\lambda)\rangle-
\int\limits_{-\infty}^{\infty}
\frac{|F^R(\mu)\rangle\langle E^L(\mu)|E^R(\lambda)\rangle}
{\mu-\lambda}\,d\mu\nonumber \\
\hspace{-7mm}&\hspace{-8mm}=&\hspace{-7mm}|E^R(\lambda)\rangle-
\int\limits_{-\infty}^{\infty}
|F^R(\mu)\rangle \tV(\mu,\lambda)\,d\mu=
|F^R(\lambda)\rangle,\\
\langle E^L(\lambda)|\hat\chi^{-1}(\lambda)&=&\langle E^L(\lambda)|+
\int\limits_{-\infty}^{\infty}
\frac{\langle E^L(\lambda)|E^R(\mu)\rangle\langle F^L(\mu)|}
{\mu-\lambda}\,d\mu
\nonumber \\
\hspace{-7mm}&\hspace{-8mm}=&\hspace{-7mm}\langle E^L(\lambda)|
-\int\limits_{-\infty}^{\infty}
\tV(\lambda,\mu)\langle F^L(\mu)| \,d\mu=
\langle F^L(\lambda)|.
\end{eqnarray}
Due to the property $\langle E^L(\lambda)| E^R(\lambda)\rangle=0$,
we find for $\lambda\in {\bf R}$
\begin{eqnarray}
\hat\chi_-(\lambda)|E^R(\lambda)\rangle&=&
\hat\chi_+(\lambda)\biggl(\hat I +2\pi i|E^R(\lambda)\rangle
\langle E^L(\lambda)|\biggr)|E^R(\lambda)\rangle \nonumber \\
&=&\hat\chi_+(\lambda)|E^R(\lambda)\rangle,\\
\langle E^L(\lambda)|\hat\chi^{-1}_-(\lambda)&=&
\langle E^L(\lambda)|\biggl(\hat I -2\pi i|E^R(\lambda)\rangle
\langle E^L(\lambda)|\biggr)\hat\chi^{-1}_+(\lambda)
\nonumber \\
&=&
\langle E^L(\lambda)|\hat\chi^{-1}_+(\lambda).
\end{eqnarray}
Here we have used
$$
\hat G^{-1}(\lambda)=\hat I -2\pi i|E^R(\lambda)\rangle\EL.
$$
The theorem is proved.

The latest theorem allows one to consider the transformation
\begin{eqnarray}\label{gauge}
\vert E^R(\lambda)\rangle \rightarrow \vert F^R(\lambda)\rangle
=\hat{\chi}(\lambda) \ERl
\end{eqnarray}
as a gauge transformation,  which is analytic in the complex plane
$\lambda \in {\bf C}$.

Thus, we have proved the equivalence of the integral equations
\Eq{intequation1p}, \Eq{intequation2p} and the \RH \Eq{eqn:inf}, \Eq{RH1}.
It is interesting to mention that integral equations  allows us to consider
the transformation $\FRl\to\ERl$ as transformation in the space
$\widetilde{\cal H}$. At the same time, the solution $\hx(\lambda)$
of the \RH defines the same transformation in the space
$\widehat{\cal H}$.
%
%%%%%%%%%%%%%%%%%%%%%%%%%%%%%%%%%%%%%%%%%%%%%%%%%%
\section{Lax representation}
%%%%%%%%%%%%%%%%%%%%%%%%%%%%%%%%%%%%%%%%%%%%%%%%%%
\par
In the previous paper \cite{KKS2} we had obtained a generalization of
the Nonlinear Schr\"o\-din\-ger equation, which describes the
\hspace{-1pt} Fredholm determinant
$\det\left(\tilde{I}+\widetilde{V}\right)$

\ba{equat1prim}
-i\partial_t\hb_{12}(u,v)&=&-\partial_x^2\hb_{12}(u,v)\non
&&\hskip-2cm
+2\stint\,dw_1dw_2\hb_{12}(u,w_1)\hb_{21}(w_1,w_2)\hb_{12}(w_2,v),\\
\label{equat2prim}
i\partial_t\hb_{21}(u,v)&=&-\partial_x^2\hb_{21}(u,v)\non
&&\hskip-2cm
+2\stint\,dw_1dw_2\hb_{21}(u,w_1)\hb_{12}(w_1,w_2)\hb_{21}(w_2,v).
\ea

In this section we derive the equations (\ref{equat1prim}),
(\ref{equat2prim}) and
their Lax representation in terms of the Riemann--Hilbert problem
(\ref{RH1}). Namely, we prove that \RH \Eq{RH1} implies equations
(\ref{equat1prim}), (\ref{equat2prim}).

In the paper \cite{KKS2} we have obtained the system of linear
differential equations for vectors $\ER$ and $\EL$. Let us
recall this system
\begin{eqnarray}
\partial_x \ER=\widehat{L}(\mu) \ER,
&~&
\partial_x \EL=-\EL \widehat{L}(\lambda),\label{Lax?1}\\
\partial_t \ER= \widehat{M}(\mu) \ER,
&~&
\partial_t \EL=-\EL \widehat{M}(\lambda),\label{Lax?2}
\end{eqnarray}
where we have set
\begin{eqnarray}
\widehat L(\lambda)=\lambda \hat{\sigma}+[\hat{g}, \hat{\sigma}],~~
\widehat M(\lambda)=-\lambda \widehat L(\lambda)+\partial_x \hat{g}.
\label{L-M1}
\end{eqnarray}
Here the operator $\hat{g}$ is defined in (\ref{defg}), the operator
$\hs$ is given by
\be{defsigma}
\hs=\frac{1}{2i}\left(\begin{array}{lr}
1&0\\0&-1\end{array}\right)\delta(u-v).
\ee
The proof of the system \Eq{Lax?1}, \Eq{Lax?2} is based on the
obvious relations
\begin{eqnarray}
\partial_x
e^{\tau(\lambda)}=-i\lambda e^{\tau(\lambda)},\qquad
\partial_t e^{\tau(\lambda)}=i\lambda^2 e^{\tau(\lambda)}.
\end{eqnarray}
It is easy to check that the system \Eq{Lax?1}, \Eq{Lax?2} is
compatible. Indeed, the compatibility condition
$$
\partial_t \widehat L-\partial_x \widehat M +
[\widehat L,\widehat M]=0
$$
implies
\be{compg}
[\partial_t\hg,\hs]-\partial^2_x\hg+\biggl[
[\hg,\hs]\cdot\partial_x\hg\biggr]=0,
\ee
which in turn is equivalent to
$$
i\partial_t \hg_{12}(u,v)-\partial^2_x \hg_{12}(u,v)=0.
$$
The last equality is obviously valid due to \Eq{defg}.

Our aim now is to find the system describing the derivatives of
vectors $\FRl$ and $\FL$ with respect to $x$ and $t$. The solution of
this problem is given by the following theorem.
\begin{thm}~~~
The vectors $|F^R(\lambda)\rangle$ and $\FL$
satisfy the following system of differential equations
\begin{eqnarray}
\partial_x \FRl= \widehat{\cal L}(\lambda) \FRl,&~&
\partial_x \FL=-\FL \widehat{\cal L}(\lambda),
\label{Lax1}\\
\partial_t \FRl= \widehat{\cal M}(\lambda) \FRl,&~&
\partial_t \FL=-\FL \widehat{\cal M}(\lambda),
\label{Lax2}
\end{eqnarray}
where we have set
\begin{eqnarray}\label{LL}
\widehat{\cal L}(\lambda)&=&
\lambda \hat{\sigma}+[\hat{b}, \hat{\sigma}],\num
\widehat{\cal M}(\lambda)&=&-\lambda \widehat{\cal L}(\lambda)
+\partial_x \hat{b}.\label{MM}
\end{eqnarray}
Here the operators $\hat{b}$ and $\hat{\sigma}$ are defined in
(\ref{defb}) and (\ref{defsigma}).
\end{thm}
{\sl Proof.}~~
It follows immediately from (\ref{gauge1}), (\ref{gauge2})
and (\ref{Lax?1}), (\ref{Lax?2})
that the derivatives of vector $\FRl$  with respect to $x$ and $t$
can be presented in the form
\begin{eqnarray}
\partial_x \FRl&=& \widehat{\cal L}(\lambda) \FRl,\\
\partial_t \FRl&=& \widehat{\cal M}(\lambda) \FRl,
\end{eqnarray}
where
\be{calLM}
{
\begin{array}{l}
{\dis \widehat{\cal L}=\partial_x\hxl\cdot
\hat\chi^{-1}(\lambda)
+\hat\chi(\lambda)\cdot \widehat L(\lambda)
\cdot\hat\chi^{-1}(\lambda),}\num
{\dis \widehat{\cal M}=\partial_t\hxl\cdot
\hat\chi^{-1}(\lambda)
+\hat\chi(\lambda)\cdot \widehat M(\lambda)
\cdot\hat\chi^{-1}(\lambda)} . \num
\end{array}
}
\ee
It was shown in the previous section that the gauge transformation
\Eq{gauge} is continuous across the real axis. The direct corollary
of this property is that matrices $\widehat{\cal L}$ and
$\widehat{\cal M}$ possess no cuts on the real axis. Let us prove
this directly.

Let $\widehat{\cal L}_\pm(\lambda)
=\lim_{\epsilon\to+0}\widehat{\cal L}(\lambda\pm i\epsilon)$. Then
we have for $\Im\lambda=0$
\be{calLpm}
\widehat{\cal L}_\pm(\lambda)=
\partial_x\hat\chi_\pm(\lambda)\cdot\hat\chi^{-1}_\pm(\lambda)
+\hat\chi_\pm(\lambda)\cdot \widehat L(\lambda)\cdot
\hat\chi_\pm^{-1}(\lambda).
\ee
Using the jump condition \Eq{RH1} we have
\ba{calLmin}
{\dis\widehat{\cal L}_-(\lambda)}&=&
{\dis \partial_x\hat\chi_-(\lambda)\cdot\hat\chi^{-1}_-(\lambda)
+\hat\chi_-(\lambda)\cdot \widehat L(\lambda)\cdot\hat
\chi_-^{-1}(\lambda)}\non
&=&{\dis
\partial_x\hat\chi_+(\lambda)\cdot\hat\chi^{-1}_+(\lambda)
+\hat\chi_+(\lambda)\cdot\partial_x\hG\cdot
\widehat G^{-1}(\lambda)\cdot\hat\chi^{-1}_+(\lambda)}\non
&+&{\dis
\hat\chi_+(\lambda)\cdot\widehat G(\lambda)\cdot \widehat L(\lambda)
\cdot\widehat G^{-1}(\lambda)\cdot \hat\chi_+^{-1}(\lambda).}
\ea
Via \Eq{Lax?1}, \Eq{Lax?2} and due to \Eq{hG}, \Eq{hH} we find
$$
\partial_x\hG=[\widehat L(\lambda),\hG].
$$
After the substitution of this expression into \Eq{calLmin} we arrive at
\be{calLplus}
\widehat{\cal L}_-(\lambda)=\partial_x\hat\chi_+(\lambda)
\cdot\hat\chi^{-1}_+(\lambda)
+\hat\chi_+(\lambda)\cdot \widehat L(\lambda)
\cdot\hat\chi_+^{-1}(\lambda)=
\widehat{\cal L}_+(\lambda).
\ee
Thus, $\widehat{\cal L}$ is continuous across the real axis. One can
perform just the same consideration for matrix $\widehat{\cal M}$ also.

Due to the analyticity properties of $\hxl$, $\hat\chi^{-1}(\lambda)$ and
their normalization conditions, we conclude that matrices
$\widehat{\cal L}(\lambda)$ and $\widehat{\cal M}(\lambda)$ are
holomorphic for $\lambda\in{\bf C}$ and possess the asymptotics
\be{asycalLM}
\widehat{\cal L}(\lambda)\to\lambda\hs+O(1),\qquad
\widehat{\cal M}(\lambda)\to-\lambda^2\hs+O(\lambda).
\ee
Due to the Liouville theorem we conclude that $\widehat{\cal
L}(\lambda)$ is a linear function of $\lambda$, while
$\widehat{\cal M}(\lambda)$ is a quadratic function of $\lambda$.
Using the asymptotic  expansion  \Eq{expansion1} we find
\ba{calLres}
&&\hspace{-7mm}{\dis
\widehat{\cal L}(\lambda)=\lambda\hat\sigma+[\hat b,\hat\sigma],}\num
&&\hspace{-7mm}{\dis\label{calMres}
\widehat{\cal M}(\lambda)=-\lambda^2\hat\sigma-\lambda[\hat
b,\hat\sigma]+\partial_x\hg
+[\hs,\hC]+[\hb,\hs]\cdot\hB-\hB\cdot[\hg,\hs]. }
\ea
The formula \Eq{calLres} exactly reproduces the
expression \Eq{LL}.  In order to reduce the expression for
$\widehat{\cal M}(\lambda)$ to the formula \Eq{MM}, one should
take into account that relations \Eq{calLM} provide us an infinite
set of identities between decomposition coefficients $\hB,~\hC,\dots$
and their derivatives with respect to $x$ and $t$. For example, the
first of the relations \Eq{calLM} implies
\be{ident1}
\partial_x\hxl=\widehat{\cal L}(\lambda)\hxl-
\hxl\widehat L(\lambda).
\ee
One can substitute  the asymptotic expansion for
$\hxl$ and explicit expressions for $\widehat{\cal L}(\lambda)$
and $\widehat L(\lambda)$ into the last formula. After this, comparing
coefficients at negative powers of $\lambda$, we obtain the  mentioned
above set of identities. In particular, for $\lambda^{-1}$ we have
\be{ident2}
\partial_x\hB=[\hs,\hC]+[\hb,\hs]\cdot\hB-\hB\cdot[\hg,\hs],
\ee
Therefore, we arrive at
\be{calMres1}
\widehat{\cal M}(\lambda)=-\lambda^2\hat\sigma-\lambda[\hat
b,\hat\sigma]+\partial_x\hb.
\ee
Thus, we have proved differential equations \Eq{Lax1}, \Eq{Lax2} for
the vector $\FRl$. The equations for vector $\FL$ can be found in
a quite similar way.

Since a gauge transformation does not disturb the compatibility
condition, we obtain for the pair (\ref{Lax1}) and (\ref{Lax2})
$$
\partial_t \widehat{\cal L}-\partial_x \widehat {\cal M}+
[\widehat{\cal L},\widehat{\cal M}]=0,
$$
what implies in turn
\be{compb}
[\partial_t\hb,\hs]-\partial^2_x\hb+\biggl[
[\hb,\hs]\cdot\partial_x\hb\biggr]=0.
\ee
The matrix operator-valued equation \Eq{compb} is equivalent to
the four scalar operator-valued partial differential equations. It is
easy to check that the equations for the diagonal part of $\hb$
are valid automatically due to identity \Eq{ident2}:
\be{diageq}
\begin{array}{l}
{\dis
\partial_x\hb_{11}(u,v)=i\stint \hb_{12}(u,w)\hb_{21}(w,v)\,dw,}\non
{\dis
\partial_x\hb_{22}(u,v)=-i\stint \hb_{21}(u,w)\hb_{12}(w,v)\,dw}
\end{array}
\ee
These equations, being substituted into the antidiagonal part of
\Eq{compb}, give the non-Abelian Nonlinear Schr\"o\-din\-ger equation
\Eq{equat1prim}, \Eq{equat2prim}. More compactly they can be written
in the following form
\be{opeq}
\begin{array}{c}
{\dis -i\partial_t\hb_{12}
=-\partial_x^2\hb_{12}+2\hb_{12}\hb_{21}\hb_{12},}\non
{\dis i\partial_t\hb_{21}
=-\partial_x^2\hb_{21}+2\hb_{21}\hb_{12}\hb_{21}.}
\end{array}
\ee
One should understand here the nonlinear terms in the sense of
integral operator products.

{\sl Remark.}~~The method of derivation of the  Nonlinear Schr\"odinger
equation described in this section is closely connected with
 the  ``dressing procedure" proposed in \cite{ZS1}.

%%%%%%%%%%%%%%%%%%%%%%%%%%%%%%%%%%%%%%%%%%%
\section{Modification of the Riemann--Hilbert problem}
%%%%%%%%%%%%%%%%%%%%%%%%%%%%%%%%%%%%%%%%%%%

In the previous sections we have demonstrated that integral
operator $\tilde I+\widetilde V$ generates in a natural way the
Riemann--Hilbert problem. The latter, in turn, defines the
dressing gauge transformation, which permits us to obtain the
exactly solvable classical differential equations. As  was shown
in the paper \cite{KKS2}, the Fredholm determinant of the
operator $\tilde I+\widetilde V$ can be described in terms of the
solutions of the \RH and differential equations mentioned. Namely,
the logarithmic derivatives of the determinant with respect to
distance and time are expressed in terms of the operator $\hx$
asymptotic expansion coefficients $\hB$ and $\hC$. Let us present
here the list of  logarithmic derivatives (see \cite{KKS2}):
\be{logdir}
\begin{array}{l}
{\dis \partial_x\log\det\freop=
i\Tr\hB_{11},}\nona{20}
{\dis\partial_t\log\det\freop=
i\Tr(\hat C_{22}-\hat C_{11}-\hB_{21}\hg_{12}),}\nona{20}
{\dis \partial_x\partial_x\log\det\freop=
-\Tr\bigl(\hb_{12}\hB_{21}\bigr),}\nona{20}
{\dis\partial_t\partial_x\log\det\freop=
i\Tr(\partial_x \hb_{12}\cdot \hB_{21}
-\partial_x \hB_{21}\cdot \hb_{12}).}
\end{array}
\ee
Thus, the solution $\hx$ of the Riemann--Hilbert problem allows to
reconstruct the Fredholm determinant up to a constant factor, which
does not depend on $x$ and $t$. Hence,  the
calculation of the correlation function of local fields is reduced to
the solving of the operator-valued Riemann-Hilbert problem.

In the present section we formulate a new Riemann--Hilbert problem,
which appears to be more convenient from the point of vie of an
asymptotic analysis.  The detailed asymptotic investigation of this
problem will be given in our forthcoming publication. Here we
restrict our selves only to basic formulations.

Consider the following substitution
\be{subst}
\hat{\chi}(\lambda)=\hmod(\lambda)\hat{\chi}^0(\lambda),
\ee
where the triangular matrix ${\chi}^0(\lambda \vert u,v)$
is defined by
\be{chi0}
\hat{\chi}^0(\lambda|u,v)=
\left(\begin{array}{cc}
1 & 0 \\
0 & 1
\end{array}\right)\delta(u-v)
+\left(\begin{array}{cc}
0 & -1\\
0 & 0
\end{array}
\right)\frac{\hg_{12}(u,v)}{u-\lambda}.
\ee
The expansion of $\hx^0$ at $\lambda\to\infty$ is given by
\be{asychi0}
\hat{\chi}^0(\lambda)=
\hat{I}-\frac{\hat{g}}{\lambda}-
\frac{i \partial_x\hat{g}}{\lambda^2}+\cdots.
\ee
Due to \Eq{expansion1} we find for the asymptotic expansion of
$\hmod$
\be{asyhmod}
\hmod(\lambda)=\hat{I}+\frac{\hat{b}}{\lambda}+
\frac{\hat{c}}{\lambda^2}+
\dots,
\ee
where the operator-valued matrix $\hb$ was defined in \Eq{defb}:
$\hb=\hB+\hg$; and
\be{hc}
\hat{c}=\hat{C}+\hat{B}\hat{g}+i\partial_x \hat{g}.
\ee

The new operator-valued matrix $\hmod(\lambda)$ satisfies the modified \rh:
\be{RHmod}
\begin{array}{c}
{\dis\hmod(\lambda)\to\hat{I},\qquad \lambda\to\infty,}\nona{20}
{\dis\hmod_-(\lambda)=\hmod_+(\lambda)\hGm(\lambda),
\qquad\lambda \in {\bf R}.}
\end{array}
\ee
The modified jump matrix $\hGm$ is equal to
\be{Gmod}
\hGm(\lambda)=\hx_+^0(\lambda)\hG
\bigl(\hx_-^0\bigr)^{-1}(\lambda),
\ee
and  possesses the following entries
\begin{eqnarray}
\hspace{-5mm}\hGm_{11}(\lambda \vert u,v)
&=&\delta(u-v)
-\delta(u-\lambda)Z(v,\lambda)\vartheta(\lambda)
e^{\phi_{D_1}(\lambda)},
\label{weak1} \\
\hspace{-5mm}\hGm_{12}(\lambda \vert u,v)
&=&-2\pi i (1-\vartheta(\lambda))\delta(u-\lambda)\delta(v-\lambda)
e^{\psi(\lambda)+\tau(\lambda)}
\label{weak2} \\
\hspace{-5mm}\hGm_{21}(\lambda \vert u,v)
&=&-\frac{i}{2\pi}\vartheta(\lambda)Z(u,\lambda)Z(v,\lambda)
e^{\phi_{D_1}(\lambda)+\phi_{A_2}(\lambda)-\psi(\lambda)-\tau(\lambda)},
\label{weak3} \\
\hspace{-5mm}\hGm_{22}(\lambda \vert u,v)
&=&\delta(u-v)
-Z(u,\lambda)\delta(v-\lambda)\vartheta(\lambda)
e^{\phi_{A_2}(\lambda)}.
\label{weak4}
\end{eqnarray}
To derive the above matrix elements, we have used the relation
$$
Z(u,u)=e^{-\phi_{D_1}(u)}+e^{-\phi_{A_2}(u)}.
$$

{\sl Remark.}~~~
One should understand the equations (\ref{weak1})--(\ref{weak4})
in the weak topology sense.
Namely,
\begin{eqnarray}
H_1(u_1,u_2)=H_2(u_1,u_2), &&(Weak~topology)
~ \Longleftrightarrow \nonumber \\
\displaystyle \int\limits_{-\infty}^{\infty}
f_1(u_1)H_1(u_1,u_2)f_2(u_2)du_1 du_2
&=&
\int\limits_{-\infty}^{\infty}
f_1(u_1)H_2(u_1,u_2)f_2(u_2)du_1 du_2.\nonumber
\end{eqnarray}
where $f_1(u_1), f_2(u_2)$ are test functions.

The transformation, considered above, is the direct generalization
of the approach proposed in \cite{IIKV} for the free fermionic
limit of the quantum Nonlinear Schr\"odinger equation.
The main advantage of the modified Riemann--Hilbert problem (\ref{RHmod})
is the simple explicit dependency of the jump matrix
$\hGm(\lambda)$ on the variables $x$ and $t$. This allows us,
in particular, to use another method for deriving the differential
equations \Eq{opeq}. Indeed, the jump condition for the matrix
\begin{eqnarray}
\hat{\chi}^{(e)}(\lambda \vert u,v)=
\hmod(\lambda \vert u,v)\exp\left(-\frac{1}{2}
\tau(\lambda)\sigma_3\right)
\end{eqnarray}
can be written as
\be{jumpconst}
\hat{\chi}_-^{(e)}(\lambda)=
\hat{\chi}_+^{(e)}(\lambda)\hGe(\lambda),\qquad
~~\lambda \in
{\bf R},
\ee
where the jump matrix
\be{jumpconst1}
\hGe(\lambda)=\exp\left(-\frac{1}{2}\tau(\lambda)\sigma_3\right)
\hGm(\lambda)\exp\left(\frac{1}{2}\tau(\lambda)\sigma_3\right),
\ee
does not depend on $x$ and $t$. Hence, the logarithmic derivatives
\begin{eqnarray}
{\cal F}_x(\lambda)=\left(\partial_x
{\hx}^{(e)}(\lambda)\right)
\bigl(\hat{\chi}^{(e)}\bigr)^{-1}(\lambda),\label{Lax1''}
\\
{\cal F}_t(\lambda)=\left(\partial_t
\hat{\chi}^{(e)}(\lambda)\right)
\bigl(\hat{\chi}^{(e)}\bigr)^{-1}(\lambda),\label{Lax2''}
\end{eqnarray}
possess no  cut on the real axis. Therefore they are holomorphic for
$\lambda\in{\bf C}$ and have the following asymptotics:
\be{logdiras}
{\cal F}_x(\lambda)\to\lambda\hs,\qquad
{\cal F}_t(\lambda)\to-\lambda^2\hs.
\ee
Due to the Liouville theorem we conclude that
${\cal F}_x(\lambda)$ and  ${\cal F}_t(\lambda)$ are linear and
quadratic functions of $\lambda$ respectively. Using the asymptotic
expansion \Eq{asyhmod} we arrive at
\begin{eqnarray} {\cal F}_x(\lambda)&=&\lambda
\hat{\sigma}+[\hat{b},\hat{\sigma}],\\
{\cal F}_t(\lambda)&=&-\lambda^2 \hat{\sigma}-\lambda[\hat{b},\hat{\sigma}]
+\partial_x\hb.
\end{eqnarray}
Thus, we again have obtained the Lax representation \Eq{LL},
\Eq{MM}.

Finally, let us rewrite the logarithmic derivatives of the Fredholm
determinant in terms on new decomposition coefficients $\hb$ and
$\hat c$. We have
\be{logdir1}
\begin{array}{l}
{\dis \partial_x\log\det\freop=
i\Tr\hb_{11}, }\nona{20}
{\dis
\partial_x\partial_x\log\det\freop=
-\Tr\bigl(\hb_{12}\hb_{21}\bigr),}\nona{20}
{\dis \partial_t\log\det\freop=
i\Tr(\hat c_{22}-\hat c_{11}),}\nona{20}
{\dis
\partial_t\partial_x\log\det\freop=
i\Tr(\partial_x \hb_{12}\cdot \hb_{21}
-\partial_x \hb_{21}\cdot \hb_{12}).}
\end{array}
\ee
%
%%%%%%%%%%%%%%%%%%
\section*{Summary}
%%%%%%%%%%%%%%%%%%
The main purpose of this paper was the formulation of the
Riemann--Hilbert problem associated with the correlation
function of the quantum Nonlinear Schr\"odinger equation.  We used the
Riemann--Hilbert problem in order to derive the non-Abelian classical
Nonlinear Schr\"odinger equation, which describes the Fredholm
determinant.  As we have seen, the solution of this equation is
completely described by the solution of the \rh. This permits us to
reduce the calculation of the Fredholm determinant to the investigation
of the \rh. The detailed asymptotic analysis of the latter   will be
performed in our forthcoming publication.

%%%%%%%%%%%%%%%%%%%%%%%%%%%%%%%%%%%%%%%%%%%%%%
\section*{Acknowledgements}
%%%%%%%%%%%%%%%%%%%%%%%%%%%%%%%%%%%%%%%%%%%%%%%
We would like to thank A.~R.~Its and T.~Kojima for useful discussions.
This work was supported by the National Science Foundation (NSF)
under Grant No. PHY-9321165,
the Russian Foundation of Basic Research under
Grant No. 96-01-00344 and INTAS-93-1038.
%%%%%%%%%%%%%%%%%%%%%%%%%%%%%%%%%%%%%%%%%%%%%%%%%%%%%%%%%%%%%%

\end{document}